\documentclass{article}

\usepackage{arxiv}
\usepackage[utf8]{inputenc} 
\usepackage[T1]{fontenc}    
\usepackage{hyperref}       
\usepackage{url}            
\usepackage{booktabs}       
\usepackage{amsfonts}       
\usepackage{nicefrac}       
\usepackage{microtype}      
\usepackage{lipsum}		
\usepackage{amsmath} 
\usepackage{graphicx}
\usepackage{natbib}
\usepackage{doi}
\usepackage{url,hyperref,lineno,microtype,subcaption}
\usepackage{comment}
\usepackage{multirow}
\usepackage{float}
\usepackage[onehalfspacing]{setspace}
\usepackage{graphicx}
\usepackage[normalem]{ulem}
\useunder{\uline}{\ul}{}
\usepackage{booktabs}
\usepackage{color}
\usepackage{xcolor}

\title{ZS-MSTM: Zero-Shot Style Transfer for Text and Speech Driven Gesture Animation using Adversarial Disentanglement of Multimodal Style Encoding}

\date{} 					

\author{ 
{
\hspace{1mm}Mireille Fares}\\
    ISIR, STMS \\
	Sorbonne University\\
	Paris, France\\
	\And
 {
 \hspace{1mm}Catherine Pelachaud} \\
    ISIR, CNRS \\
	Sorbonne University\\
	Paris, France\\
 \And
  {
  \hspace{1mm}Nicolas Obin}\\
    STMS \\
	Sorbonne University\\
	Paris, France\\}

\renewcommand{\headeright}{ }
\renewcommand{\undertitle}{ }
\renewcommand{\shorttitle}{Zero-Shot Style Transfer for Gesture Animation}

\hypersetup{
pdftitle={A template for the arxiv style},
pdfsubject={q-bio.NC, q-bio.QM},
pdfauthor={David S.~Hippocampus, Elias D.~Striatum},
pdfkeywords={First keyword, Second keyword, More},
}

\begin{document}
\maketitle

\begin{abstract}
In this study, we address the importance of modeling behavior style in virtual agents for personalized human-agent interaction. We propose a machine learning approach to synthesize gestures, driven by prosodic features and text, in the style of different speakers, even those unseen during training. Our model incorporates zero-shot multimodal style transfer using multimodal data from the PATS database, which contains videos of diverse speakers. We recognize style as a pervasive element during speech, influencing the expressivity of communicative behaviors, while content is conveyed through multimodal signals and text. By disentangling content and style, we directly infer the style embedding, even for speakers not included in the training phase, without the need for additional training or fine-tuning. Objective and subjective evaluations are conducted to validate our approach and compare it against two baseline methods.
\end{abstract}


\section{Introduction}
\textcolor{black}{\textit{Embodied Conversational Agents} are virtually embodied agents with a human-like appearance that are capable of autonomously communicating with people in a socially intelligent manner using multimodal behaviors (\cite{lugrin2021introduction}). The field of research in \textit{ECAs} has emerged as new interface between humans and machines. ECAs behaviors are often modeled from human communicative behaviors. They are endowed with the capacities to recognize and generate verbal and non-verbal cues (\cite{lugrin2021introduction}), and are envisioned to support humans in their daily lives. Our work revolves around modeling multimodal data and learning the complex correlations between the different modalities employed in human communication. More specifically, the objective is to model the multimodal ECAs' behavior with their\textit{ behavior style}.}

Human \textit{behavior style} is a socially meaningful clustering of features found within and across multiple modalities, specifically in \textit{linguistic} (\cite{campbell2006elements}), \textit{spoken behavior} such as the speaking style conveyed by speech prosody (\cite{moon2022mist, Obin_2011}), and \textit{nonverbal behavior} such as hand gestures and body posture (\cite{obermeier2015speaker, wagner2014gesture}). 

\textit{Behavior style} involves the ways in which people talk differently in different situations. A same person may have different speaking styles depending on the situation (e.g. at home, at the office or with friends). These situations  can carry different social meanings (\cite{bell1984language}). Different persons may also have different behavior styles while communicating in  similar contexts. \textit{Behavior style} is syntagmatic. It unfolds over time in the course of an interaction and during one’s life course (\cite{campbell2006elements}). It does not emerge unaltered from the speaker. It is continuously attuned, accomplished and co-produced with the audience (\cite{mendoza1999style}). It can be very self-conscious and at the same time can be extremely routinized to the extent that it resists attempts of being altered (\cite{mendoza1999style}). Movements and gestures are person-specific and \emph{idiosyncratic} in nature (\cite{mcneill2005gesture}), and each speaker has his or her own non-verbal behavior style that is linked to his/her personality, role, culture, etc.

A large number of generative models were proposed in the past few years for synthesizing gestures of ECAs. Style modelling and control in gesture is receiving attention in order to propose more expressive ECAs behaviors that could possibly be adapted to a specific audience (\cite{neff2008gesture, karras2017audio, cudeiro2019capture, ahuja2020style, Ginosar_2019_CVPR, alexanderson2020style, Ahuja_CVPR_lowRes}). \textcolor{black}{They assume that \textit{behavior style} is encoded in the \textit{body gesturing}}. Some of these works generate full body gesture animation driven by text in the style of one specific speaker (\cite{neff2008gesture}). Other approaches (\cite{alexanderson2020style, karras2017audio, cudeiro2019capture, Ginosar_2019_CVPR}) are speech-driven. For some of these approaches, the behavior style of the synthesized gestures is changed by exerting direct control over the synthesized gestures' velocity and force (\cite{alexanderson2020style}). For others (\cite{cudeiro2019capture, karras2017audio, Ginosar_2019_CVPR}), they produce the gestures in the style of a \emph{single speaker} by training their generative models on one \emph{single speaker}'s data, and synthesizing the gestures corresponding to this specific speaker's audio. Moreover, verbal and non-verbal behavior plays a crucial role in communication in human-human interaction (\cite{36}). Generative models that aim to predict communicative gestures of ECAs must produce expressive semantically-aware gestures that are aligned with speech (\cite{cassell00nudge}).  

We propose a novel approach to model \textit{behavior style} in ECAs and to tackle the different \textit{behavior style} modeling challenges. We view \textit{behavior style} as being pervasive while speaking; it colors the communicative behaviors expressivity while speech content is carried by multimodal signals and text. To design our approach, we make the following assumptions for the separation of style and content information: \emph{style} is possibly encoded across all modalities (text, speech, pose) and varies little or not over time; \emph{content} is encoded only by text and speech modalities and varies over time. Our approach aims at (1) synthesizing natural and expressive upper body gestures of a source speaker, by encoding the \emph{content} of two input modalities – text semantics and Mel spectrogram, (2) conditioning the source speaker’s predicted gesture on the multimodal \emph{style} representation of a target speaker, and therefore rendering the model able to perform style transfer across speakers, and finally (3) allowing zero-shot style transfer of newly coming speakers that were not seen by the model during training. The disentanglement scheme of \textit{content} and \textit{style} allows us to directly infer the style embedding even of speakers whose data are not part of the training phase, without requiring any further training or fine-tuning.

Our model consists of two main components: first (1) a speaker style encoder network which goal is to model a specific target speaker style extracted from three input modalities – Mel spectrogram, upper-body gestures, and text semantics; and second (2) a sequence-to-sequence synthesis network that generates a sequence of upper-body gestures based on the content of two input modalities – Mel spectrogram and text semantics – of a source speaker, and conditioned on the target speaker style embedding. Our model is trained on the \textit{multi-speaker} database PATS, which was proposed in \cite{ahuja2020style} and designed to study gesture generation and style transfer. It includes 3 main modalities that we are considering in our approach: text semantics represented by BERT embeddings, Mel spectrogram and 2D upper body poses.

Our contributions can be listed as follows:
\begin{enumerate}
    \item We propose the first approach for zero-shot multimodal style transfer approach for 2D pose synthesis. At inference, an embedding style vector can be directly inferred from multimodal data (text, speech and poses) of any speaker, by simple projection into the embedding style space (similar to the one used in \cite{jia2018transfer}). The style transfer performed by our model allows the transfer of style from any unseen speakers, without further training or fine-tuning of our trained model. Thus it is not limited to the styles of the speakers of a given database.    
    \item Unlike the work of \cite{ahuja2020style} and previous works, the encoding of the style takes into account 3 modalities: body poses, text semantics, and speech - Mel spectrograms; which are important for gesture generation (\cite{kucherenko2019analyzing, Ginosar_2019_CVPR}) and linked to style. We encode and disentangle \emph{content} and \emph{style} information from multiple modalities. On one side, a content encoder is used to encode a content matrix from text and speech signal; on the other hand, a style encoder is used to encode a style vector from all text, speech, and pose modalities. A fader loss is introduced to effectively disentangle content and style encodings (\cite{lample2017fader}).
\end{enumerate}
\textcolor{black}{In the following sections, we first discuss the related works and more specifically the existing behavior style modelling approaches, as well as their limitations. Next, in Section \ref{materials}, we dive into the details of our model's architecture, describe its training regime, and the objective and subjective evaluations we conducted. We then discuss in Section \ref{results} the objective and subjective evaluation results. Next, in Section \ref{discussion}, we review the key findings of our work, compare it to prior research and discuss its main limitations. We conclude by discussing future directions for our work. }

\section{Related Work}
Since few years, a large number of gesture generative models have been proposed,  principally based on sequential generative parametric models such as Hidden Markov Models HMM and gradually moving towards deep neural networks enabling spectacular advances over the last few years. Hidden Markov Models were previously used to predict head motion driven by prosody (\cite{sargin2008analysis}), and body motion (\cite{levine2009real, marsella2013towards}). 
 
\cite{chiu2014gesture} proposed an approach for predicting gesture labels from speech using conditional random fields (CRFs) and generating gesture motion based on these labels, using Gaussian process latent variable models (GPLVMs). These works focus on the gesture generation task driven by either one modality namely speech, or by the two modalities - speech and text. Their work focuses on producing naturalistic and coherent gestures that are aligned with speech and text, enabling a smoother interaction with ECAs, and leveraging the vocal and visual prosody. The non-verbal behavior is therefore generated in conjunction with the verbal behavior.   
LSTM networks driven by speech were recently used to predict sequences of gestures (\cite{hasegawa2018evaluation}) and body motions (\cite{shlizerman2018audio, ahuja2019react}). LSTMs were additionally employed for synthesizing sequences of facial gestures driven by text and speech, namely the fundamental frequency (F0) (\cite{fares2020towards, fares2021multimodalwacai}). Generative adversarial networks (GANs) were  proposed to generate realistic head motion (\cite{sadoughi2018novel}) and body motions (\cite{ferstl2019multi}). Furthermore, transformer networks and attention mechanisms were recently used for upper-facial gesture synthesis based on multimodal data - text and speech (\cite{fares2021multimodal}). 
\cite{jonell2020let} propose a probabilistic approach based on normalizing flows for synthesizing facial gestures in dyadic settings.
\textcolor{black}{Facial (\cite{fares2021multimodal, fares2020towards}) and hand (\cite{kucherenko2020gesticulator})} gestures driven by both acoustic and semantic information are the closest approaches to our gesture generation task, however they cannot be used for the style transfer task.

Beyond realistic generation of human non-verbal behavior, style modelling and control in gesture is receiving more attention in order to propose more expressive behaviors that could possibly adapted to a specific audience (\cite{neff2008gesture, karras2017audio, cudeiro2019capture, ahuja2020style, Ginosar_2019_CVPR, alexanderson2020style, Ahuja_CVPR_lowRes}).  
\cite{neff2008gesture} propose a system that produces full body gesture animation driven by text, in the style of a specific performer.  
\cite{alexanderson2020style} propose a generative model for synthesizing speech-driven gesticulation, they exert directorial control over the output style such as gesture level and speed.  
\cite{karras2017audio} propose a model for driving 3D facial animation from audio. Their main objective is to model the style of a single actor by using a deep neural network that outputs 3D vertex positions of meshes that correspond to a specific audio.  
\cite{cudeiro2019capture} also propose a model that synthesizes 3D facial animation driven by speech signal. 
\textcolor{black}{The learned model, VOCA (Voice Operated Character Animation) takes any speech signal as input--even speech in languages other than English--and realistically animates a wide range of adult faces. Conditioning on subject labels during training allows the model to learn a variety of realistic speaking styles. VOCA also provides animator controls to alter speaking style, identity-dependent facial shape, and pose (i.e. head, jaw, and eyeball rotations) during animation. }

\cite{Ginosar_2019_CVPR} propose an approach for generating gestures given audio speech, however their approach uses models trained on single speakers. The aforementioned works have focused on generating nonverbal behaviors (facial expression, head movement, gestures in particular) aligned with speech (\cite{neff2008gesture, karras2017audio, cudeiro2019capture, ahuja2020style}). They have not consider multimodal data when modeling style, as well as when synthesizing gestures.  

To our knowledge, the only attempts to model and transfer the style from multi-speakers database have been proposed by \cite{ahuja2020style} and \cite{Ahuja_CVPR_lowRes}. \cite{ahuja2020style} presented Mix-StAGE, a speech driven approach that trains a model from multiple speakers while learning a unique style embedding for each speaker. They created PATS, a dataset designed to study various styles of gestures for a large number of speakers in diverse settings. 
In their proposed neural architecture, a content and a style encoder are used to extract content and style information from speech and pose. To disentangle style from content information, they assume that style is only encoded through the pose modality, and the content is shared across speech and pose modalities. A style embedding matrix whose each vector represents the style associated to a specific speaker from the training set.
During training, they further propose a multimodal GAN strategy to generate poses either from the speech or pose modality. During inference, the pose is inferred by only using the speech modality and the desired style token. 
 
However, their generative model is conditioned on gesture style and driven by audio. It does not include verbal information. It cannot perform zero-shot style transfer on speakers that were not seen by their model during training. In addition, the style is associated with each unique speaker, which makes the distinction unclear between each speaker's specific style - idiosyncrasy -, the style that is shared among a set of speakers of similar settings (i.e. TV show hosts, journalists, etc...), and the style that is unique to each speaker's prototype gestures that are produced consciously and unconsciously. Moreover, the style transfer is limited to the styles of PATS speakers, which prevents the transfer of style from an unseen speaker. Furthermore, the proposed architecture is based on the disentangling of content and PATS style information, which is based on the assumption that style is only encoded by gestures. However, both text and speech also convey style information, and the encoding of style must take into account all the modalities of human behavior. To tackle those issues, \cite{Ahuja_CVPR_lowRes} presented a few-shot style transfer strategy based on neural domain adaptation accounting for cross-modal grounding shift between source speaker and target style. This adaptation still requires 2 \textcolor{black}{minutes} of the style to be transferred. 
To the best of our knowledge, our approach is the first to synthesize  gestures from a source speaker, which are semantically-aware, speech driven and conditioned on a multimodal representation of the style of target speakers, in a zero-shot configuration i.e., without requiring any further training or fine-tuning. 

\section{Materials and Methods}\label{materials}
\subsection{Model Architecture}
We propose \textbf{ZS-MSTM} (\textbf{Z}ero-\textbf{S}hot \textbf{M}ultimodal \textbf{S}tyle \textbf{T}ransfer \textbf{M}odel), a Transformer-based architecture for stylized upper-body gesture synthesis, driven by the content of a source speaker's speech - text semantics represented by BERT embeddings and audio Mel spectrogram -, and conditioned on a target speaker's multimodal style embedding. The stylized generated gestures correspond to the style of target speakers that have been seen and unseen during training. 
\begin{figure}[h!]
\centering
\includegraphics[width=\textwidth]{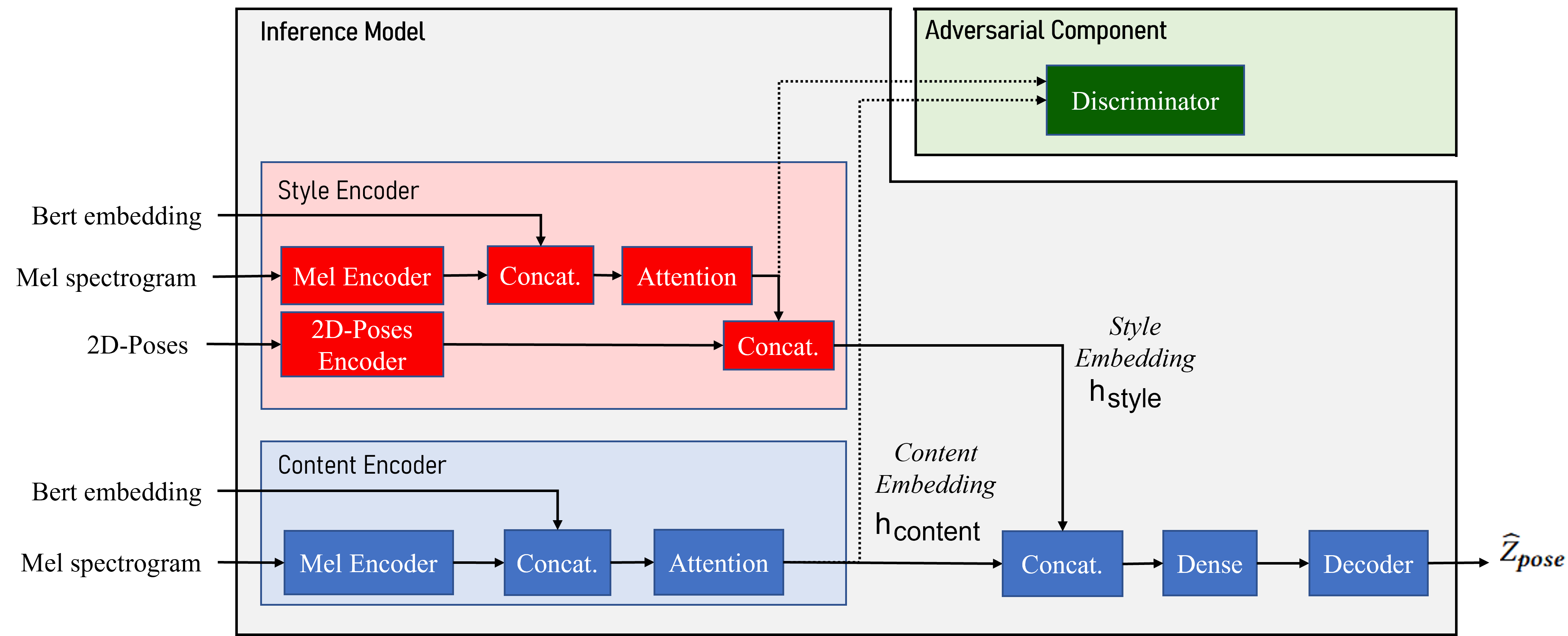}
\caption{ \textbf{ZS-MSTM} (\textbf{Z}ero-\textbf{S}hot \textbf{M}ultimodal \textbf{S}tyle \textbf{T}ransfer \textbf{M}odel) architecture. The content encoder (further referred to as $E_{content}$) is used to encode content embedding $h_{content}$ from  BERT text embeddings $X_{text}$ and speech Mel-spectrograms $X_{speech}$ using a speech encoder $E_{speech}^{content}$. The style encoder (further referred to as $E_{style}$) is used to encode style embedding $h_{style}$ from multimodal text $X_{text}$, speech $X_{speech}$, and pose $X_{pose}$ using speech encoder  $E_{speech}^{style}$ and pose encoder $E_{pose}^{style}$. The generator $G$ is a transformer network that generates the sequence of poses $\widehat{Z}_{pose}$ from the sequence of content embedding $h_{content}$ and the style embedding vector $h_{style}$. The adversarial module relying on the discriminator $Dis$ is used to disentangle content and style embeddings $h_{content}$ and $h_{style}$. }
\label{fig:ArchitectureOverview}
\end{figure}

As depicted in Figure \ref{fig:ArchitectureOverview}, the system is composed of three main components:
\begin{enumerate}
    \item A \textbf{speaker style encoder} network that learns to generate a fixed-dimensional speaker embedding style from a \emph{target speaker} multimodal data: 2D poses, BERT embeddings, and Mel spectrogram, all extracted from videos in a database.
    \item A \textbf{sequence to sequence gesture synthesis} network that synthesizes upper-body behavior (including hand gestures and body poses) based on the content of two input modalities - text embeddings and Mel spectrogram - of a \emph{source speaker}, and conditioned on the \emph{target speaker} style embedding. A \emph{content encoder} is presented to encode the content of the Mel spectrogram along with BERT embeddings.
    \item An \textbf{adversarial component} in the form of a fader network (\cite{lample2017fader}) is used for disentangling style and content from the multimodal data. 
\end{enumerate}

At inference time, the adversarial component is discarded, and the model can generate different versions of poses when fed with different style embeddings.   
Gesture styles for the same input speech can be directly controlled by switching the value of the style embedding vector $h_{style}$ or by calculating this embedding from a target speaker's multimodal data fed as input to the \emph{Style Encoder}.
 
ZS-MSTM illustrated in Fig. \ref{fig:ArchitectureOverview} aims at mapping multimodal speech and text feature sequences into continuous upper-body gestures, conditioned on a speaker style embedding. The network operates on a segment-level of 64 timesteps: the inputs and output of the network consist of one feature vector for each segment \emph{\textbf{S}} of the input text sequence. The length of the segment-level input features (text and audio) corresponds to $t$ = $64$ timesteps (as provided by \emph{PATS Corpus}). The model generates a sequence of gestures corresponding to the same segment-level features given as inputs. Gestures are sequences of 2D poses represented by \emph{x} and \emph{y} positions of the joints of the skeleton. The network has an embedding dimension $d_{model}$ equal to $768$. 

\subsubsection{Content Encoder}
The content encoder $E_{content}$ illustrated in Figure \ref{fig:ArchitectureOverview} takes as inputs BERT embedding $X_{text}$ and audio Mel spectrograms $X_{speech}$ corresponding to each \textbf{S}.  $X_{text}$ is represented by a vector of length 768 - BERT embedding size used in \textit{PATS Corpus}. $X_{speech}$ is encoded using \emph{Mel Spectrogram Transformer (AST)} pre-trained \emph{base384} model (\cite{gong2021ast}). 
 
\emph{AST} operates as follows: the input Mel spectrogram which has $128$ frequency bins, is split into a sequence of 16x16 patches with overlap, and then is linearly projected into a sequence of 1D patch vectors, which is added with a positional embedding. We append a [\emph{CLS}] token to the resulting sequence, which is then input to a \emph{Transformer Encoder}. \emph{AST} was originally proposed for audio classification. Since we do not intend to use it for a classification task, we remove the linear layer with sigmoid activation function at the output of the \emph{Transformer Encoder}. We use the \emph{Transformer Encoder}'s output of the [\emph{CLS}] token as the Mel spectrogram representation \textbf{S}. The \emph{Transformer Encoder} has an embedding dimension equals to $d_{model}$, $N_{enc}$ equals to 12 encoding layers, and $N_{h}$ equals to 12 attention heads.

The segment-level encoded Mel spectrogram is then concatenated with the segment-level BERT embedding. A self-attention mechanism is then applied on the resulting vector. The multi-head attention layer has $N_{h}$ equals to 4 attention heads, and an embedding size $d_{att}$ equals to $d_{att}=d_{model}+768$. The output of the attention layer is the vector $h_{content}$, a content representation of the source speaker's segment-level Mel spectrogram and text embedding, and it can be written as follows:
\begin{equation} \label{eqn1}
	h_{content} =  sa\left( \left[ E_{speech}^{content} (X_{speech}),  X_{text} \right] \right)
\end{equation}
where: sa(.) denotes self-attention.

\subsubsection{Style Encoder}
As discussed previously, \emph{behavior style} is a clustering of features found within and across modalities, encompassing verbal and non-verbal behavior. It is not limited to gestural information. We consider that \emph{behavior style} is encoded in a speaker's multimodal - text, speech and pose - behavior. As illustrated in Figure \ref{fig:ArchitectureOverview}, the style encoder $E_{style}$  takes as input, at the segment-level,  Mel spectrogram $X_{speech}$, BERT embedding $X_{text}$, and a sequence of (X, Y) joints positions that correspond to a target speaker's 2D poses $X_{pose}$. \emph{AST} is used to encode the audio input spectrogram. $N_{lay}$ equals to 3 layers of LSTMs with a hidden-size equal to $d_{model}$ are used to encode the vector representing the 2D poses. The last hidden layer is then concatenated with the audio representation. Next, a multi-head attention mechanism is  applied on the resulting vector. This attention layer has $N_{h}$ equals to 4 attention heads and an embedding size equals to $d_{att}$. Finally, the output vector is  concatenated with the 2D poses vector representation. The resulting vector $h_{style}$  is the output speaker style embedding that serves to condition the network with the speaker style. The final style embedding $h_{style}$ can therefore be written as follows:
\begin{equation} \label{eqn2}
		h_{style} = \left[ sa \left( \left[X_{text}, E_{speech}^{style}  (X_{speech}) \right] \right),  E_{pose}^{style} (X_{pose})   \right] 
\end{equation}
where: sa(.) denotes self-attention.
\subsubsection{Sequence to sequence gesture synthesis}
The stylized 2D poses are generated given the sequence of content representation $h_{content}$ of the source speaker's  Mel spectrogram and text embeddings obtained at $S$-level, and conditioned by the style vector embedding $h_{style}$   generated from a target speaker's multimodal data. For decoding the stylized 2D-poses, the sequence of $h_{content}$ and the vector $h_{style}$ are concatenated (by repeating the $h_{style}$ vector for each segment of the sequence), and passed through a $Dense$ layer of size $d_{model}$.  
We then give the resulting vector as input to a \textit{Transformer Decoder}. The \textit{Transformer Decoder} is composed of $N_{dec}$\emph{ = 1} decoding layer, with $N_{h}$ = 2 attention heads, and an  embedding size equal to $d_{model}$. Similar to the one proposed in \cite{vaswani2017attention}, it is composed of residual connections applied around each of the sub-layers, followed by layer normalization. Moreover, the self-attention sub-layer in the decoder stack is altered  to prevent positions from attending to subsequent positions. The output predictions are offset by one position. This masking makes sure that the predictions for position index \emph{\textbf{j}} depends only on the known outputs at positions that are less than \emph{\textbf{j}}. For the last step, we  perform a permutation of the first and the second dimensions of the vector generated by the transformer decoder. The resulting vector is a sequence of 2D-poses which corresponds to: 
\begin{equation}\label{loss_zsmstm1.0pose}
\widehat{Z}_{pose} = G(h_{content}, h_{style})
\end{equation}
where: G is the transformer generator conditioned on latent content embedding $h_{content}$ and style embedding $h_{style}$ .  
The generator loss of the transformer gesture synthesis can be written as:
\begin{equation} \label{eqn3}
	\mathcal{L}^{gen}_{rec}(E_{content}, E_{style}, G) = \mathbb{E}_{\widehat{Z}_{pose}} ||\widehat{Z}_{pose} - G(h_{content}, h_{style}) ||_2
\end{equation}
\subsubsection{Adversarial Component}
\begin{figure}[]
\includegraphics[width=12cm]{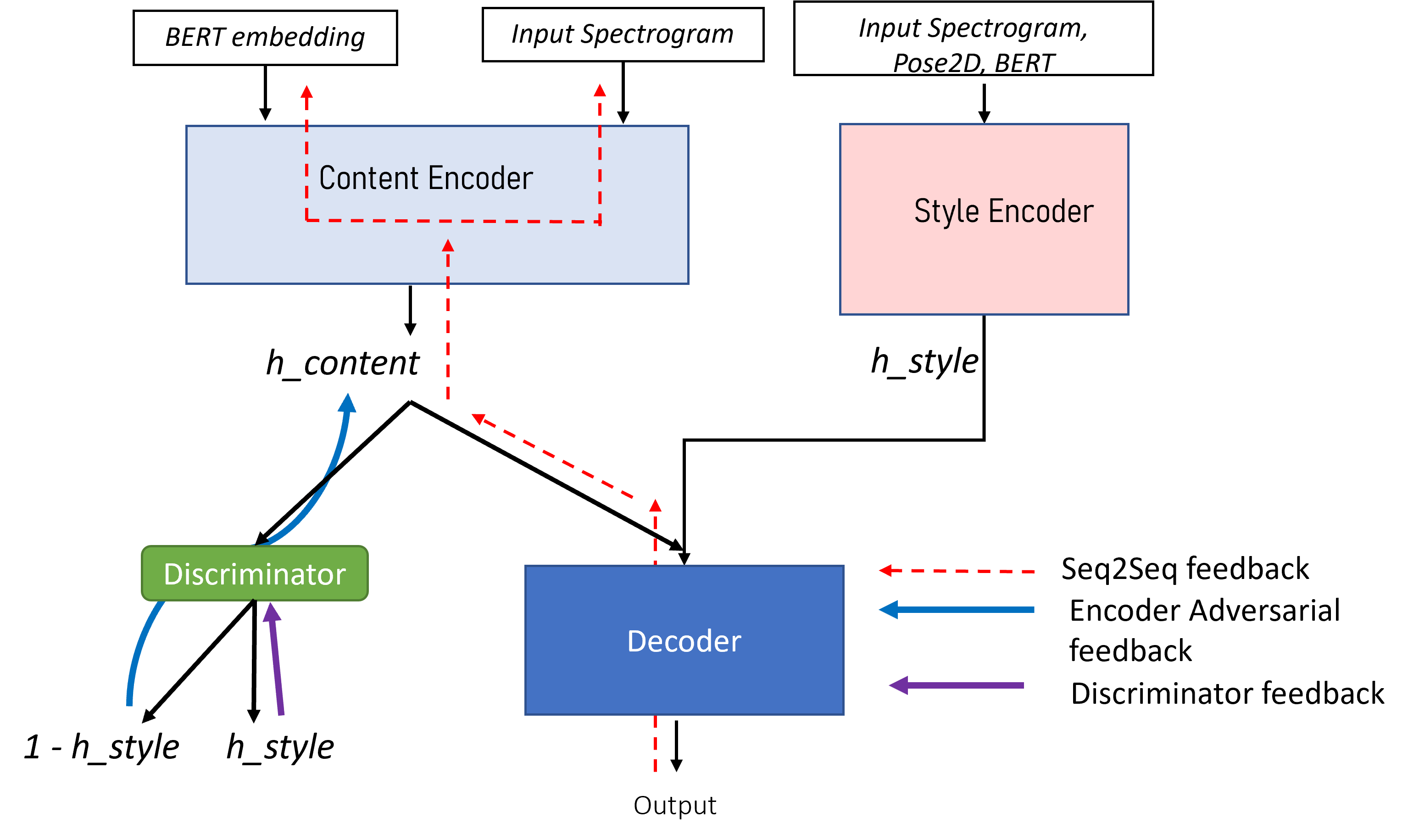}
\centering
\caption{Fader network for multimodal content and style disentangling.}
\label{fig:FaderNetwork}
\end{figure}

Our approach of disentangling style from content relies on the fader network disentangling approach (\cite{lample2017fader}), where a fader loss is introduced to effectively separate content and style encodings, as depicted in Figure \ref{fig:FaderNetwork}. The fundamental feature of our disentangling scheme is to constrain the latent space of $h_{content}$ to be independent of the style embeddings $h_{style}$. Concretely, it means that the distribution over $h_{content}$ of the latent representations should not contain the style information. A fader network is composed of: an encoder which encodes the input information \emph{X} into the latent code $h_{content}$, a decoder which decodes the original data from the latent, and an additional variable $h_{style}$ used to condition the decoder with the desired information (a face attribute in the original paper). The objective of the fader network is to learn a latent encoding $h_{content}$ of the input data that is independent on the conditioning variable $h_{style}$ while both variables are complementary to reconstruct the original input data from the latent variable $h_{content}$ and the conditioning variable $h_{style}$. To do so, a discriminator \emph{Dis} is optimized to predict the variable $h_{style}$ from the latent code $h_{content}$; on the other side the auto-encoder is optimized using an additional adversarial loss so that the classifier \emph{Dis} is unable to predict the variable $h_{style}$. Contrary to the original fader network in which the conditional variable is discrete within a finite binary set (0 or 1 for the presence or absence attribute), in this paper the conditional variable $h_{style}$ is continuous. We then formulate this discriminator as a regression on the conditional variable $h_{style}$: the discriminator learns to predict the style embedding $h_{style}$ from the content embedding $h_{content}$, as:
\begin{equation}
\widehat{h}_{style} = Dis({h}_{content})
\end{equation}
While optimizing the discriminator, the discriminator loss $\mathcal{L}^{dis}$ must be as low as possible, such as:
\begin{equation}
\mathcal{L}^{dis}(D) = \mathbb{E}_{\widehat{h}_{style}} ||{h}_{style} - Dis(h_{content}) ||_2
\end{equation}
 
In turn, optimizing the generator loss including the fader loss $\mathcal{L}^{gen}_{adv}$, the discriminator must not be able to predict correctly the style embedding $h_{style}$ from the content embedding $h_{content}$ conducting to a high discriminator error and thus a low fader loss. 
The adversarial loss can be written as,
\begin{equation}
\mathcal{L}^{gen}_{adv}(E_{content}, E_{style}, G) = \mathbb{E}_{\widehat{h}_{style}} || 1 - (h_{style} - Dis(h_{content})) ||_2
\end{equation}
To be consistent, the style prediction error is preliminary normalized within 0 and 1 range.
 
\noindent Finally, the total generator loss can therefore be written as follows:
\begin{equation} \label{eqn4}
	\mathcal{L}^{gen}_{total}(E_{content}, E_{style}, G)  = \mathcal{L}^{gen}_{rec}(E_{content}, E_{style}, G) + \lambda \mathcal{L}^{gen}_{adv}(E_{content}, E_{style}, G)
\end{equation}
where $\lambda$ is the adversarial weight that starts off at 0 and is linearly incremented by 0.01 after each training step. 

The discriminator $Dis$ and the generator $G$ are then optimized alternatively as described in \cite{lample2017fader}.

All \textbf{ZS-MSTM} hyperparameters were chosen empirically and are summarized in Table \ref{ZS_MSTM_hyperparameters}.

\begin{table}[H]
\centering
\begin{tabular}{llll}
\hline
\textbf{Component}                              & \multicolumn{2}{c}{\textbf{Hyperparameter}} & \textbf{Value} \\ \hline \hline
\multirow{3}{*}{AST (base384 model)}            & Embedding size          & $d_{model}$       & 768            \\ \cline{2-4} 
                                 & Encoding layers & $N_{lay}$ & 12   \\ \cline{2-4} 
                                 & Attention heads & $N_{h}$  & 12   \\ \hline
\multirow{2}{*}{Content Encoder} & Attention heads & $N_{h}$  & 4    \\ \cline{2-4} 
                                 & Embedding size  & $d_{att}$    & 1536 \\ \hline
\multirow{3}{*}{Style Encoder}   & \multirow{2}{*} {2D Pose LSTMs}   & $N_{lay}$ & 3    \\ \cline{3-4} 
                                                                   & & $N_{hid}$ & 768\\ \cline{2-4} 
                                 & Attention heads & $N_{h}$  & 4    \\ \cline{2-4} 
                                 & Embedding size  & $d_{att}$    & 1536 \\ \hline
\multirow{3}{*}{Sequence to Sequence Component} & Transformer Decoder     & $N_{dec}$     & 1              \\ \cline{2-4} 
                                 & Attention heads & $N_{h}$  & 2    \\ \cline{2-4} 
                                 & Embedding size  & $d_{model}$  & 768  \\ \hline
\end{tabular}
\caption{ZS-MSTM hyperparameters}
\label{ZS_MSTM_hyperparameters}
\end{table}

\subsection{Training Regime}
This section describes the training regime we follow for training \textbf{ZS-MSTM}. 
We trained our network using the \textit{PATS Corpus} (\cite{ahuja2020style}). PATS was created to study various styles of gestures. The dataset contains upper-body 2D pose sequences aligned with corresponding Mel spectrogram, and BERT embeddings. It offers 251 hours of data, with a mean of 10.7 seconds and a standard deviation of 13.5 seconds per interval. PATS gathers data from 25 speakers with different behavior styles from various settings (e.g., lecturers, TV shows hosts). It contains also several annotations. The spoken text has been transcribed in PATS and aligned with the speech. The 2D body poses have been extracted with OpenPose.  

Each speaker is represented by their lexical diversity and the spatial extend of their arms. While in PATS arms and fingers have been extracted, we do not consider finger data in our work. That is we do not model and predict 2D finger joints. This choice arises as the analysis of finger data is very noisy and not very accurate. We model 11 joints that represent upper body and arm joints. 

We consider two test conditions: \emph{Seen Speaker} and \emph{Unseen Speaker}. The \emph{Seen Speaker} condition aims to assess the style transfer correctness that our model can achieve when presented with speakers that were seen during training as target style. On the other hand, the \emph{Unseen Speaker} condition aims to assess the performance of our model when presented with unseen target speakers, to perform zero-shot style transfer. Seen and unseen speakers are specifically selected from PATS to cover a diversity of stylistic behavior with respect to lexical diversity and spatial extent as reported by \cite{ahuja2020style}.

For each PATS speaker, there is a train, validation and test set already defined in the database. For testing the \emph{Seen Speaker} condition, our test set includes the train sets of 16 PATS speakers. Six other speakers are selected for the \emph{Unseen Speaker} condition, and their test sets are also used for our experiments. These six speakers differ in their behavior style and lexical diversity. \textit{Seen} and \textit{Unseen} speakers are listed in Table \ref{Seen_and_Unseen_PATS_Speakers}.
\begin{table}[H]
\centering
\begin{tabular}{ll}
\hline
\textbf{Condition}                          & \multicolumn{1}{c}{\textbf{Speakers}}                               \\ \hline \hline
\multicolumn{1}{c}{Seen}                    & "Shelly", "Jon", "Fallon", "Bee", "Ellen",                          \\
                                            & "Oliver", "Lec\_cosmic", "Lec\_hist", "Ytch\_prof", "Ytch\_dating", \\
                                            & "Seth", "Conan", "Angelica", "Rock", "Noah",  and "Lec\_law"        \\ \hline
\multicolumn{1}{c}{\multirow{2}{*}{Unseen}} & "Lec\_evol", "Almaram", "Huckabee",                                 \\
\multicolumn{1}{c}{}                        & "Ytch\_charisma", "Minhaj", and "Chemistry                          \\ \hline
\end{tabular}
\caption{Seen and Unseen PATS Speakers}
\label{Seen_and_Unseen_PATS_Speakers}
\end{table}

\textcolor{black}{We developed our model using Pytorch and trained it on an NVIDIA Corporation GP102 (GeForce GTX 1080 Ti) machine. Each training batch contains $BS$ = 24 pairs of word embeddings, Mel spectrogram, and their corresponding sequence of (X, Y) joints of the skeleton (of the upper-body pose). We use Adam optimizer with $\beta_{1} = 0.95, \beta_{2} = 0.999$. For balanced learning, we use a scheduler with an initial learning rate $Lr$ equals to 1e-5, with $W_{steps}$ equals to $20,000$. We train the network for $N_{ep}$ = 200. All features values are normalized so that the dataset mean and standard deviation are 0 and 0.5, respectively. Table \ref{TrainingHyperparameters} summarizes all hyperparameters used for training.}

\begin{table}[H]
\centering
\begin{tabular}{lll}
\hline
\multicolumn{2}{c}{\textbf{Hyperparameter}}                            & \textbf{Value} \\ \hline \hline
Batch Size                      & $BS$                                 & 24             \\ \hline
Number of epochs                & $N_{ep}$                          & 200            \\ \hline
\multirow{2}{*}{Adam Optimizer} & $\beta_{1}$                          & 0.95           \\ \cline{2-3} 
                                & $\beta_{2}$                          & 0.999          \\ \hline
\multirow{2}{*}{Scheduler}      &  $W_{steps}$ & 20,000         \\ \cline{2-3} 
                                & $Lr$                                 & 1e-5           \\ \hline
\end{tabular}
\caption{Training Hyperparameters}
\label{TrainingHyperparameters}
\end{table}

\subsection{Objective Evaluation}
To validate our approach and assess the stylized generated gestures, we conduct an objective evaluation for the two conditions \textit{Seen Speakers} and \textit{Unseen Speakers}. 

\subsubsection{Objective Metrics}
In our work, we have defined \textit{behavior style} by the \textit{behavior expressivity} of a speaker. To evaluate objectively our works, we define metrics to compare the \textit{behavior expressivity} generated by our model, with the target speaker's \textit{behavior expressivity}, and source speaker's \textit{behavior expressivity}.

Following works on \textit{behavior expressivity} by \cite{Wallbott98} and \cite{pelachaud2009studies}, we define 4 objective \textit{behavior dynamics} metrics to evaluate the style transfer of different target speakers:  \textit{acceleration}, \textit{jerk} and \textit{velocity} that are averaged over the values of all upper-body joints, as well as the speaker's average \textit{bounding box perimeter} (BB perimeter) of his/her body movements extension.

In addition, we compute the \textit{acceleration}, \textit{jerk} and \textit{velocity} of only the \textit{left} and \textit{right} \textit{wrists}, to obtain information on the \textit{arms movements expressivity} (\cite{Wallbott98,kucherenko2019analyzing}).

For both conditions $SD$ and $SI$, we define two sets of distances: 
\begin{enumerate}
    \item \textbf{\emph{Dist.(}}\emph{Source}, \emph{Target}\textbf{)}: representing the average distance between the source style and the target style,
    \item \textbf{\emph{Dist.(}}\emph{ZS-MSTM}, \emph{Target}\textbf{)}: representing the average distance between our model's gestures style and the target style.
\end{enumerate}

More specifically, after computing the \textit{behavior expressivity} and \textit{BB perimeter} of our model's generated gestures, the ones of source speakers, and the ones of the target speakers, we calculate the average distance as follows:
\begin{equation}
\text{$Dist_{avg}$\textit{(x, Target)}} =  \frac{\text{\textbf{Dist.}\textit{(x, Target)}}}{\text{\textbf{Dist.}\textit{(Source, Target)}} + \text{\textbf{Dist.}\textit{(ZS-MSTM, Target)}}} \times 100
\end{equation}
Where x denotes \textit{Source} for computing $\textbf{Dist}_{avg} (Source, Target)$ and \textit{ZS-MSTM} for computing $\textbf{Dist}_{avg} (\textit{ZS-MSTM}, Target)$.

\textcolor{black}{
To investigate the impact of each input modality on our style encoder, we conducted ablation studies on different versions of our model. Specifically, we performed ablations of the \textit{pose modality}, \textit{text modality}, and \textit{audio modality}. We also compared the performance of the full model with that of the baseline \textit{DiffGAN} \cite{Ahuja_CVPR_lowRes}. We employ two metrics to evaluate the correlation and timing between gestures and spoken language: \textbf{Probability of Correct Keypoints (PCK)} and \textbf{L1 distance}. For PCK, we averaged the values over $\alpha$ = $0.1$ and $0.2$, as suggested in \cite{ginosar2019learning}. L1 distance was calculated between the generated gestures and the corresponding target ground truth gestures.}

\subsection{Human Perceptual Studies}
We conduct three human perceptual studies.

\begin{enumerate}
    \item \textbf{Study 1} - To investigate human perception of the stylized upper-body gestures produced by our model, we conduct a human perceptual study that aims to assess the style transfer of speakers \textit{seen} during training - \emph{Seen Speaker} condition. 
    \item \textbf{Study 2} - We conduct another human perceptual study that aims to assess the style transfer of speakers \textit{unseen} during training - \emph{Unseen Speaker} condition. 
    \item \textbf{Study 3} - We additionally conduct a third human perceptual study to compare \textbf{ZS-MSTM}'s produced stylized gestures in \emph{Seen Speaker} and \emph{Unseen Speaker} conditions, to \textit{Mix-StAGE} which we consider our baseline.
\end{enumerate}

The evaluation studies are conducted with 35 participants that were recruited through the online crowd-sourcing website Prolific. Participants are selected such that they are fluent in English. Attention checks are added in the beginning and the middle of each study to filter out inattentive participants. All the animations presented in these studies are in the form of 2D sticks.

\begin{figure} [H]
\includegraphics[width=10cm]{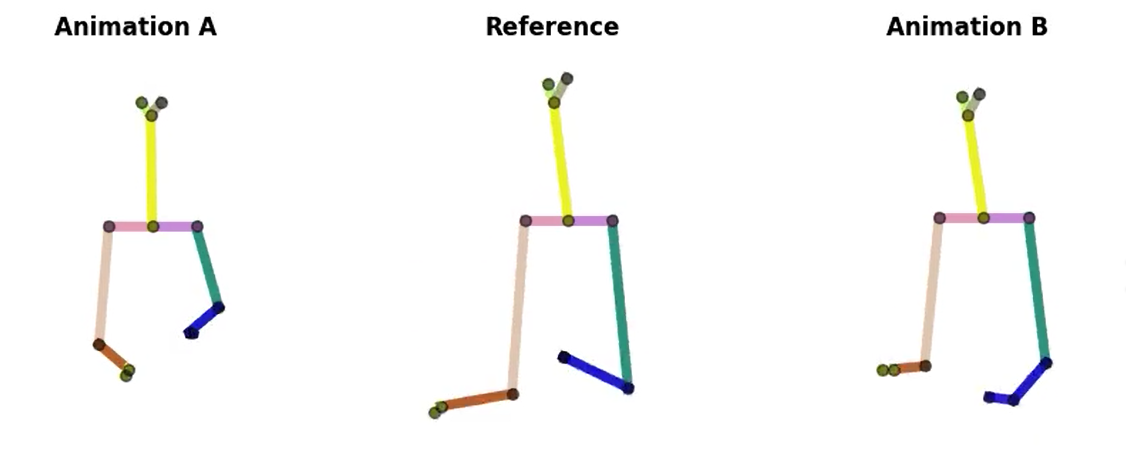}
\centering
\caption{Three 2D stick animations: \textit{Animation A}, the \textit{Reference}, and \textit{Animation B}. The target style is represented by \textit{Reference}. \textbf{ZS-MSTM}'s predictions, and the \textbf{source style} are illustrated in Animation A or B.}
\label{triplets}
\end{figure}

\textbf{Study 1 and 2. }For Study 1 and 2, we presented 60 stimuli of 2D stick animations. Each study included 30 stimuli. A stimulus is a triplet of 2D animations composed of: 
\begin{itemize}
    \item A 2D animation with the \textit{source style},
    \item A 2D animation with the \textit{target style},
    \item A 2D animation of \textit{ZS-MSTM}'s prediction after performing the style transfer.
\end{itemize}
Figure \ref{triplets} illustrates the three animations we present for each set of questions. The animation of the target style is the \textbf{\textit{Reference}}. The animation of our model's predictions, and the source style is either \textbf{\textit{Animation A}} or \textbf{\textit{Animation B}} (randomly chosen).
 
\begin{figure} []
\includegraphics[width=9cm]{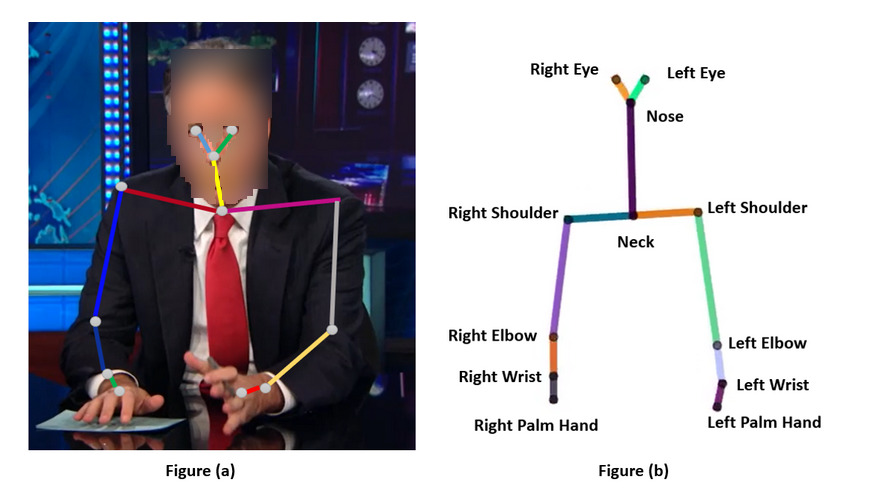}
\centering
\caption{Upper-body 2D skeleton of a speaker Vs. a virtual agent}
\label{training}
\end{figure}

For each triplet of animations, we asked 6 questions to evaluate 6 factors related to the \textit{resemblance} of the produced gestures w.r.t the the \textit{source style} and \textit{target style}: 
\begin{enumerate}
    \item Please rate \textbf{the overall resemblance of the Reference} w.r.t A and B. (\textbf{Factor 1} - Overall resemblance) 
    \item Please rate the \textbf{resemblance of the Left (L) and Right (R) arms gesturing of the Reference} w.r.t the left and right arm gesturing of A and B. (\textbf{Factor 2} - Arms gesturing)
    \item Please rate the \textbf{resemblance of the body orientation of the Reference} w.r.t the body orientation of A and B. (\textbf{Factor 3} - Body orientation)
    \item Please rate the \textbf{resemblance of the gesture amplitude of the Reference} w.r.t the gesture amplitude of A and B. (\textbf{Factor 4} - Gesture amplitude)
    \item Please rate the \textbf{resemblance of the gesture frequency of the Reference} w.r.t the gesture frequency of A and B. (\textbf{Factor 5} - Gesture frequency)
    \item Please rate the \textbf{resemblance of the gesture velocity of Reference} w.r.t the gesture velocity of A and B. (\textbf{Factor 6} - Gesture velocity)
\end{enumerate}
Each factor is rated on a \textit{5 likert} scale, as follows:
\begin{enumerate}
    \item Reference is very similar to A
    \item Reference is mostly similar to A
    \item Reference is in between A and B
    \item Reference is mostly similar to B
    \item Reference is very similar to B
\end{enumerate}
\textbf{Training. }Each study includes a training at its beginning. The training provides an overview of the 2D upper-body skeleton of the virtual agent, its composition, and gesturing. The goal of the training is to get the participants familiarized with the 2D skeleton before starting the study. More specifically, the training included a description of how the motion of a speaker in a video is extracted by detecting his/her facial and body motion and extracting his/her 2D skeleton of joints, and stated that in a similar fashion, the eyes and upper-body movement of a virtual agent are represented by a 2D skeleton of joints, as depicted in Figure \ref{training} 
Moreover, we present and describe different shots of the 2D skeleton gesturing with \textit{its right/left arms}, and with different \textit{body orientation}, which is described as the orientation of the shoulders and neck.

\textbf{Pre-tests. }We conducted pre-tests to make sure that the 2D animations are comprehensible by participants, as well as the questions. Participants reported that the training, stimuli and questions are coherent and comprehensible, however each study was too long, as it lasted 30 minutes. For this reason, we divided each study to three, such that each study includes only 10 stimuli, and is conducted by different participants. Hence, 6 studies including a pre-training, and the evaluation of 10 stimuli were conducted by 35 participants that are different.

\textbf{Study 3. }For Study 3, we present 20 stimuli consisting of triplets of 2D stick animations. Similar to \textit{Study 1} and \textit{Study 2}, for each triplet, we present: \textit{Animation A}, the \textit{Reference}, and \textit{Animation B}. The animation of the target style is the \textit{Reference}. The animation of Mix-StAGE’s predictions, and the source style is either Animation A or Animation B (randomly chosen). We note that these stimuli include the same \textit{source} and \textit{target} styles that were used in \textit{Study 1} and \textit{Study 2}, and which were randomly chosen. Study 3 also included training at its beginning, which is the same as the one previously described.

\section{Results}\label{results}
\subsection{Objective Evaluation Results}
Objective evaluation experiments are conducted for evaluating the performance of our model in the \emph{Seen Speaker} and \emph{Unseen Speaker} conditions. For \emph{Seen Speaker} condition, experiments are conducted on the test set that includes the 16 speakers that are seen by our model during training. For \emph{Unseen Speaker} condition, experiments are also conducted on another test set that includes the 6 speakers that were not seen during training.
\begin{figure}[H]
\includegraphics[width=9cm]{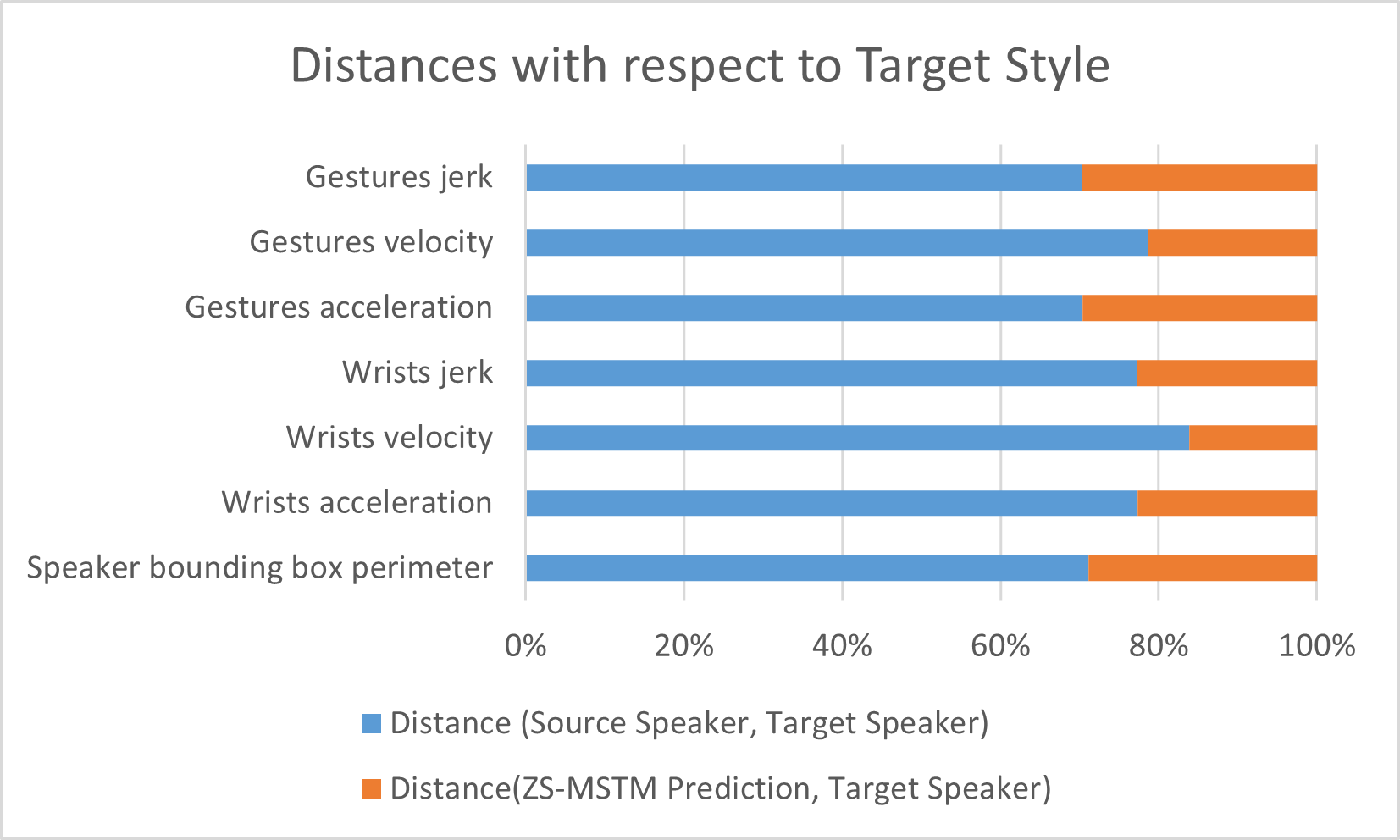}
\centering
\caption{Distances between the target speaker style and each of the source style and our model's generated gestures style for seen target speakers}
\label{dist_seen}
\end{figure}
Figure \ref{dist_seen} reports the experimental results on the \emph{Seen Speaker} test set. It illustrates the results of \textbf{\emph{Dist.(}}\emph{Source}, \emph{Target}\textbf{)} in terms of \textit{behaviors dynamics} and speaker \textit{bounding box perimeter} between the target speaker style and the source speaker style.

For \emph{Seen Speaker} condition (Figure \ref{dist_seen}), \textbf{\emph{Dist.(}}\emph{Source}, \emph{Target}\textbf{)} is higher than 70\% of the total distance for all behavior dynamics metrics.; thus \textbf{\emph{Dist.(}}\emph{ZS-MSTM}, \emph{Target}\textbf{)} is less than 30\% of the total distance for all behavior dynamics metrics. Wrists velocity, jerk and acceleration results reveal that the virtual agent's arms movements show the same expressivity dynamics as the target style (\textbf{\emph{Dist.(}}\emph{ZS-MSTM}, \emph{Target}\textbf{)} $<$ 22\%).

The style transfer from target speaker "Shelly" to source speaker "Angelica" - knowing that Angelica is a \emph{Seen Speaker} - shows that the distance of predicted gestures' behavior dynamics metrics are close (distance $<$ 20\%) to "Shelly" (\emph{target style}), while the ones between "Angelica" and "Shelly" are far (distance $>$ 80\%).

The perimeter of the prediction's bounding box (BB) is closer (distance $<$ 30 \%) to the target speaker's BB perimeter than the source . The closeness between predictions dynamics behavior metrics values are shown for all speakers in the \emph{Seen Speaker} condition, specifically for the following style transfers - \emph{target} to \emph{source} - : "Fallon" to "Shelly", "Bee" to "Shelly", "Conan" to "Angelica", "Oliver" to "lec\_cosmic", which are considered having different lexical diversity, as well as spatial average extent, as reported by the authors of PATS (\cite{ahuja2020style}).  
\begin{figure} [H]
\includegraphics[width=9cm]{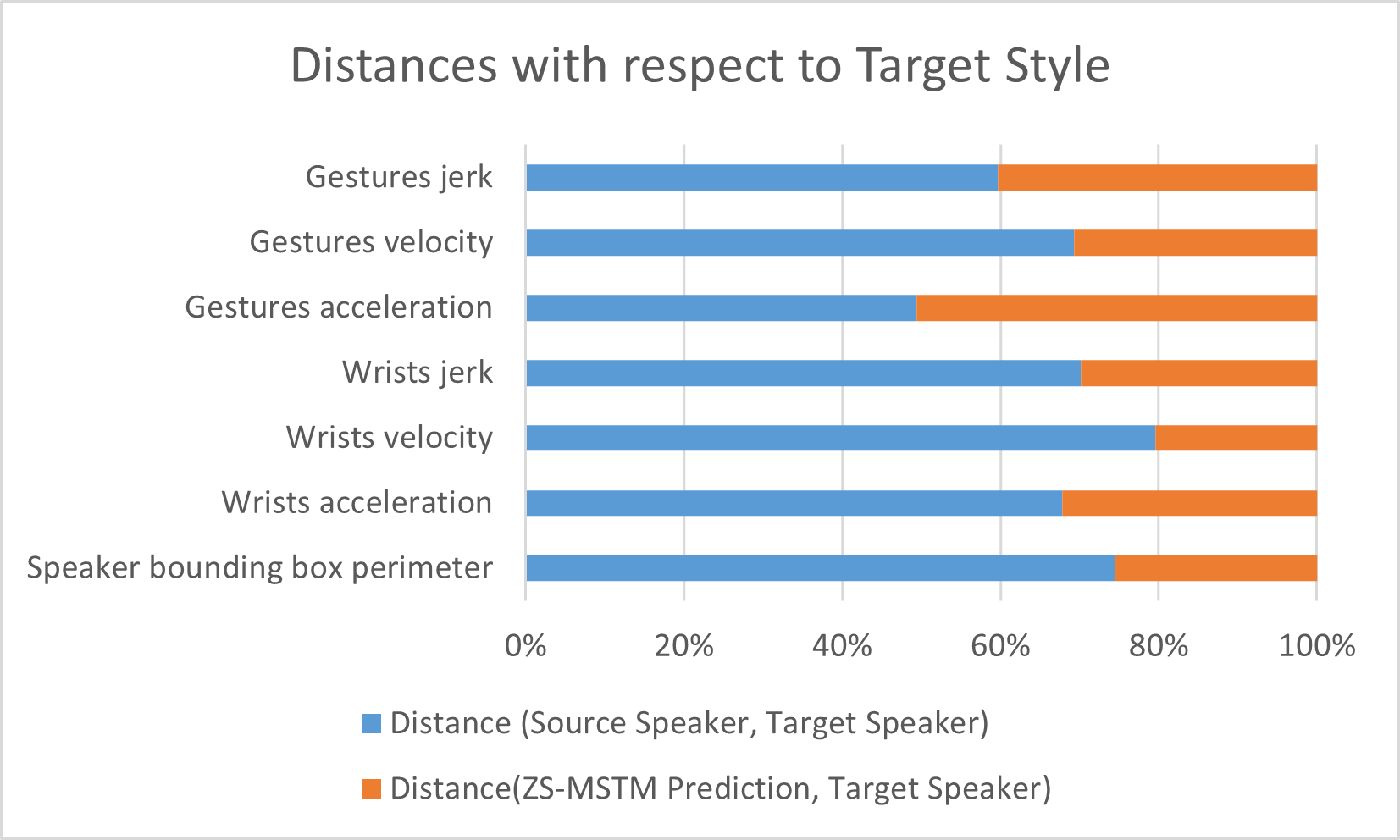}
\centering
\caption{Distances between the target speaker style and each of the source style and our model's generated gestures style for unseen target speakers}
\label{dist_unseen}
\end{figure}
Experimental results for the \emph{Unseen Speaker} test set are depicted in Fig. \ref{dist_unseen}. Results reveal that our model is capable of reproducing the style of the 6 unseen speakers. As depicted in Fig. \ref{dist_unseen}, for all behavior dynamics metrics, as well as the bounding box perimeter, \textbf{\emph{Dist.(}}\emph{Source}, \emph{Target}\textbf{)} is higher than 50\% of the total distances for all metrics. Results show that for wrists velocity, jerk and acceleration, \textbf{\emph{Dist.(}}\emph{ZS-MSTM}, \emph{Target}\textbf{)} is less than 33\%. Thus, arm movement's expressivity produced by \emph{ZS-MSTM} is close to the one of the target speaker style. Moreover, the perimeter of the prediction's bounding box is close (distance $<$ 30 \%) to the target speaker's, while the distance between the BB perimeter of the source and the target is far (distance $>$ 70 \%). While our model has not seen "Lec\_evol"'s multimodal data during training, it is yet capable of transferring his behavior expressivity style to the source speaker "Oliver".  It is also capable of performing zero-shot style transfer from the target speaker "Minhaj" to the source speaker "Conan". In fact, results show that wrists acceleration and jerk values of our model's generated gestures are very close to those of the target speaker "Minhaj". We observe the same results for the 6 speakers for the \emph{Unseen Speaker} condition. 

We additionally conduct a Fisher's LSD Test to do pair-wise comparisons on all metrics, for the two set of distances - \textbf{\emph{Dist.(}}\emph{Source}, \emph{Target}\textbf{)}, and \textbf{\emph{Dist.(}}\emph{ZS-MSTM}, \emph{Target}\textbf{)} - in both conditions. We find significant results ($p< 0.003$) for all distances in both conditions. 
\begin{table}[H]
\centering
\caption{\textcolor{black}{Comparison of our ZS-MSTM model with DiffGAN \cite{Ahuja_CVPR_lowRes} used as a baseline, as well as with different versions of our model where we removed the Text, Audio, and Pose modalities from the Style Encoder. Please note that we report here the values for \textit{DiffGAN} from \cite{Ahuja_CVPR_lowRes}}}.
\vspace{0.3cm}
\label{tab:ablation}
\resizebox{\textwidth}{!}{%
\begin{tabular}{@{}llllll|llll@{}}
\cmidrule(l){3-10}
 &
   &
  \multicolumn{4}{c|}{\textbf{L1}} &
  \multicolumn{4}{c}{\textbf{PCK}} \\ \hline\midrule
\multicolumn{1}{c}{\textbf{Models   |}} &
  \textit{\textbf{\begin{tabular}[c]{@{}l@{}}Source:\\ Target:\end{tabular}}} &
  \textbf{\begin{tabular}[c]{@{}l@{}}Oliver \\ Chemistry\end{tabular}} &
  \textbf{\begin{tabular}[c]{@{}l@{}}Maher  \\ Chemistry\end{tabular}} &
  \textbf{\begin{tabular}[c]{@{}l@{}}Oliver \\ Maher\end{tabular}} &
  \textbf{\begin{tabular}[c]{@{}l@{}}Maher \\ Oliver\end{tabular}} &
  \textbf{\begin{tabular}[c]{@{}l@{}}Oliver   \\ Chemistry\end{tabular}} &
  \textbf{\begin{tabular}[c]{@{}l@{}}Maher \\ Chemistry\end{tabular}} &
  \textbf{\begin{tabular}[c]{@{}l@{}}Oliver  \\ Maher\end{tabular}} &
  \textbf{\begin{tabular}[c]{@{}l@{}}Maher  \\ Oliver\end{tabular}} \\ \midrule
\multicolumn{2}{l}{\textbf{ZS-MSTM - Text Ablation}} &
  0.51 ± 0.05 &
  0.58 ± 0.07 &
  0.56 ± 0.07 &
  0.36 ± 0.08 &
  0.89 ± 0.78 &
  0.95 ± 0.98 &
  0.87 ± 0.89 &
  0.97 ± 0.76 \\ \midrule
\multicolumn{2}{l}{\textbf{ZS-MSTM -  Audio Ablation}} &
  0.65 ± 0.08 &
  0.71 ± 0.08 &
  0.91 ± 0.07 &
  0.89 ± 0.08 &
  0.85 ± 0.78 &
  0.82 ± 0.98 &
  0.84 ± 0.89 &
  0.95 ± 0.76 \\ \midrule
\multicolumn{2}{l}{\textbf{ZS-MSTM - Pose Ablation}} &
  0.87 ± 0.08 &
  0.91 ± 0.08 &
  1.11 ± 0.12 &
  0.76 ± 0.08 &
  0.81 ± 0.78 &
  0.9 ± 0.98 &
  0.82 ± 0.89 &
  0.92 ± 0.76 \\ \midrule
\multicolumn{2}{l}{\textbf{DiffGAN} (\cite{Ahuja_CVPR_lowRes})} &
  1.36 ± 0.03 &
  0.88 ± 0.03 &
  1.48 ± 0.01 &
  0.53 ± 0.02 &
  0.29 ± 0.01 &
  0.31 ± 0.01 &
  0.26 ± 0.01 &
  0.45 ± 0.02 \\ \midrule
\multicolumn{2}{l}{\textbf{ZS-MSTM - Full Model}} &
  \textbf{0.34 ± 0.04} &
  \textbf{0.36 ± 0.04} &
  \textbf{0.49 ± 0.05} &
  \textbf{0.11 ± 0.03} &
  \textbf{0.96 ± 0.91} &
  \textbf{0.96 ± 0.99} &
  \textbf{0.89 ± 0.92} &
  \textbf{0.97 ± 0.98} \\ \bottomrule
\end{tabular}%
}
\end{table}

\textcolor{black}{The results of our ablation studies are summarized in Table \ref{tab:ablation}. Specifically, we trained three versions of our \textit{ZS-MSTM} model, each with one modality (either text, audio, or pose) removed from the style encoder. We evaluated the performance of each model using the \textit{L1 distance} and \textit{PCK} metrics, comparing the predictions to the target ground truth in all conditions.
Our results (see Table \ref{tab:ablation}) show that the \textit{L1 distance} between the predictions of the ablated models and the ground truth is higher compared to the full model condition, for both seen (\textit{Oliver}) and unseen (\textit{Chemistry, Maher}) target styles. This trend was observed across all three ablation conditions. In addition, we compared our results to the baseline \textit{DiffGAN} (\cite{Ahuja_CVPR_lowRes}) and found that our \textbf{\textit{ZS-MSTM}} model consistently outperforms \textit{DiffGAN} in terms of \textit{L1 distance}, with higher confidence intervals reported as standard deviation on all source-target pairs.
Furthermore, we evaluated the \textbf{\textit{PCK}} metric for all source-target pairs, and found that our \textbf{\textit{ZS-MSTM}} model achieves higher accuracy than the ablated models for all style transfers, with higher confidence intervals. This indicates that our model produces joint positions that are accurate and closely match the ground truth. When comparing \textbf{\textit{ZS-MSTM}} with \textit{DiffGAN}, our model outperforms \textit{DiffGAN} in terms of \textit{PCK}, with higher confidence intervals.
}
\subsubsection{Additional t-SNE Analysis}  
In this work, the style encoder is agnostic: it is the attention weights that make it possible to exploit the different modalities given as input to the style encoder.
\begin{figure}[H]
\includegraphics[width=\textwidth]{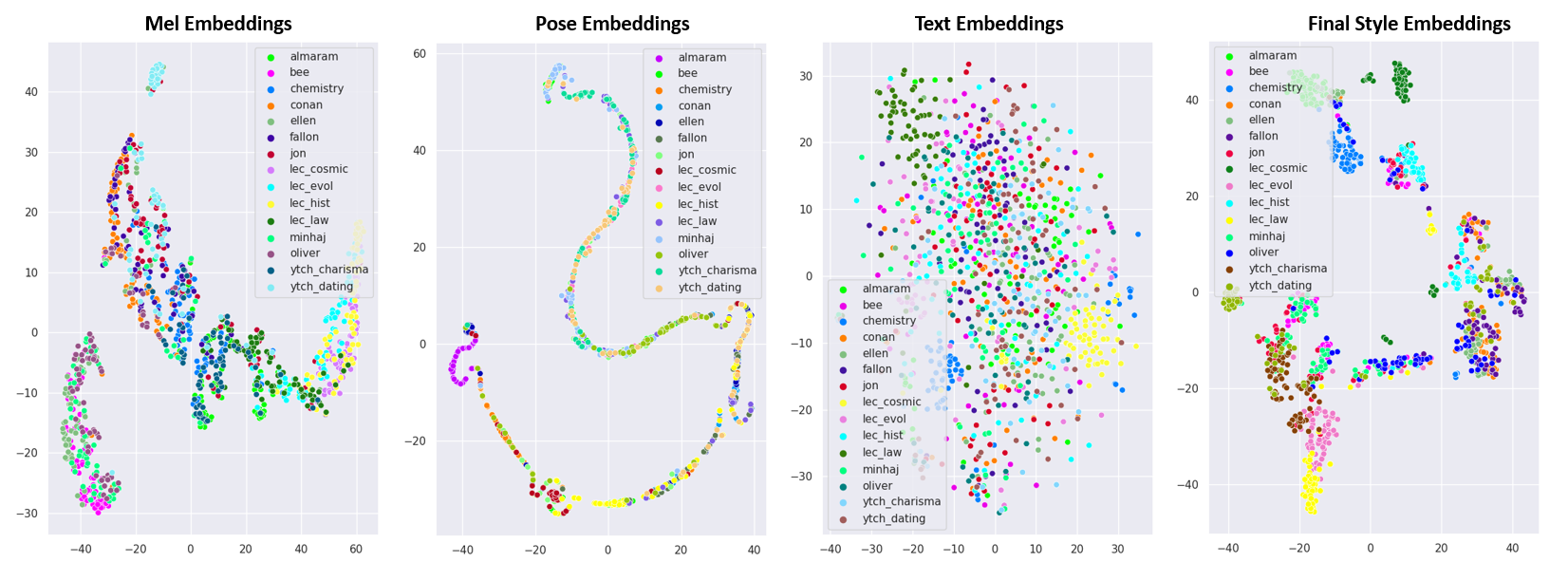}
\centering
\caption{2D TSNE Analysis of the generated \textbf{\textit{Mel Embeddings}}, \textbf{\textit{Pose Embeddings}}, \textbf{\textit{Text Embeddings}}, and the final \textbf{\textit{Style Embeddings}}}
\label{2DTSNE_Analysis}
\end{figure}
We conducted a t-SNE post-hoc analysis of the distributions of the style vectors at the output of each modality. Figure \ref{2DTSNE_Analysis} illustrates the 2D t-SNE plots of \textbf{\textit{Mel Embeddings}}, \textbf{\textit{Pose Embeddings}}, \textbf{\textit{Text Embeddings}}, and the final \textbf{\textit{Style Embeddings}} produced by our model \textbf{ZS-MSTM}. We found that the motion style depends most on the \textit{pose modality}, followed by the \textit{speech}, then the \textit{text semantics}. 

\subsection{Human Perceptual Studies Results}\label{results_perceptivestudies_chap7}
\begin{figure} [H]
\includegraphics[scale=0.5]{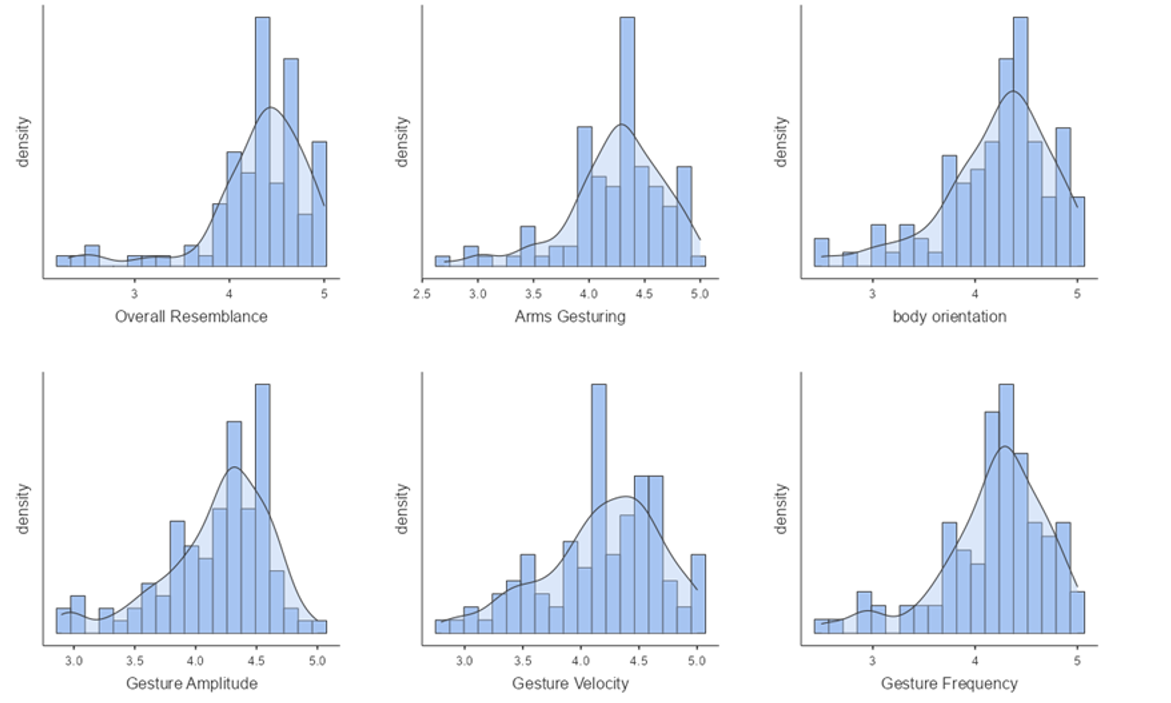}
\centering
\caption{Density plots of \textbf{\textit{Overall Resemblance}}, \textbf{\textit{Arms Gesturing}}, \textbf{\textit{Body Orientation}}, \textbf{\textit{Gesture Amplitude}}, \textbf{\textit{Gesture Frequency}}, \textbf{\textit{Gesture Velocity}} for the \textit{Seen Speakers} condition}
\label{6factors_seen}
\end{figure}

\textbf{Study 1 - \textit{Seen Speakers}. }
\begin{figure} [H]
\includegraphics[width=11cm]{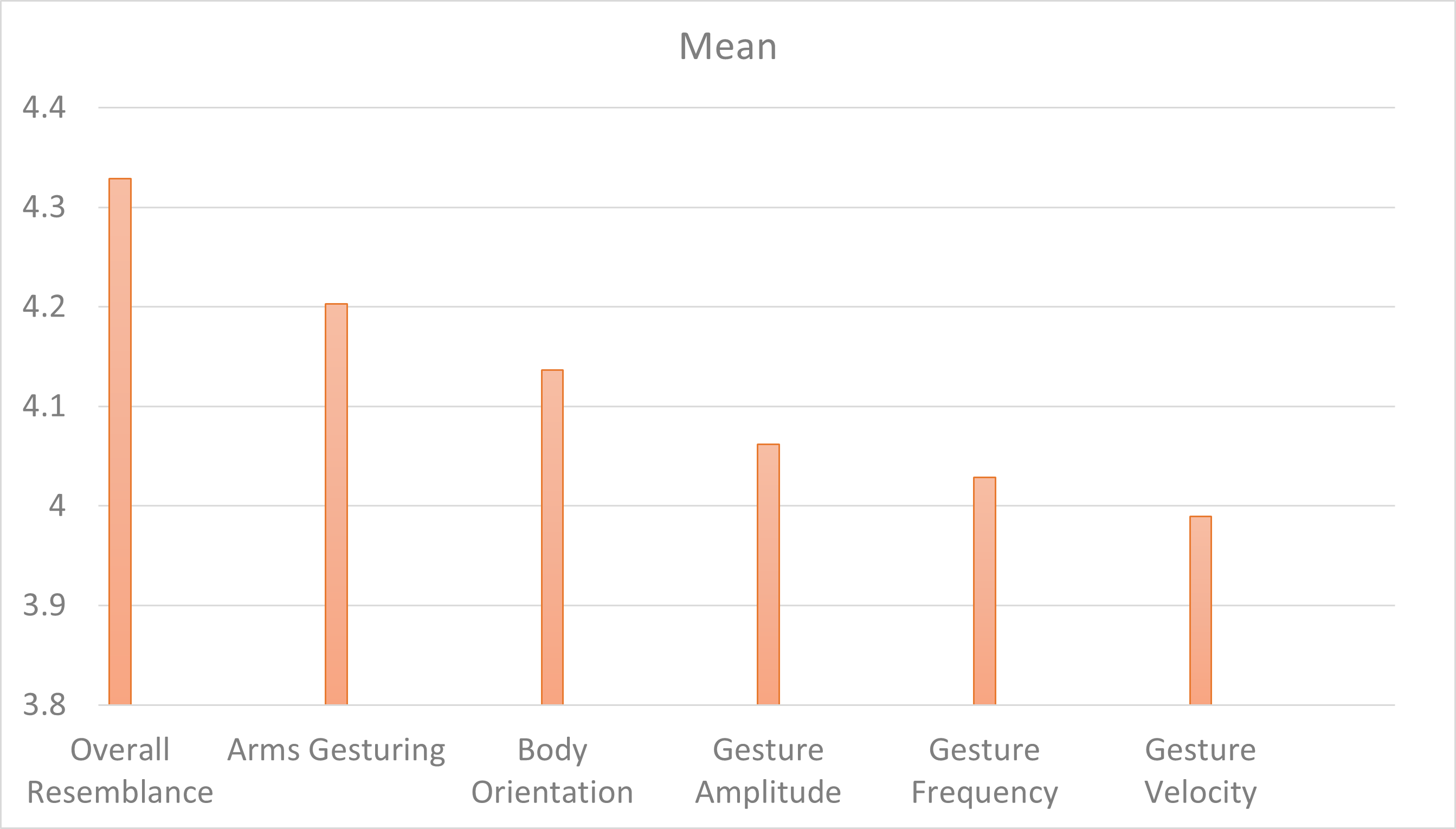}
\centering
\caption{The mean scores of all the factors for \textit{Seen Speakers} condition}
\label{results_seen}
\end{figure}
Our first perceptive study (Study 1) aims to evaluate the style transfer of speakers \textit{seen} during training. Figure \ref{results_seen} shows the mean scores obtained on the 6 factors for the condition \textit{"seen speakers"}. On a \textit{5 likert scale}, the \textbf{\textit{overall resemblance}} factor obtained a score of 4.32, which means that the \textbf{ZS-MSTM}'s 2D animations closely resemble  the 2D animations of the \textit{seen target style}. The resemblance is also reflected by the mean scores of \textbf{\textit{arms gesturing}}, \textbf{\textit{body orientation}}, \textbf{\textit{gesture amplitude}}, \textbf{\textit{gesture frequency}}, as well as \textbf{\textit{gesture velocity}}, which is between 3.99 and 4.2. We observed that for all factors, most of the participants gave a score between 3.8 and 5, as depicted in Figure \ref{6factors_seen}.

We additionally performed post-hoc paired samples t-tests between all the factors, and found significant results between \textbf{\textit{overall resemblance}} and all the other factors (p $\leq$ 0.008). 

\begin{figure} 
\includegraphics[width=13cm]{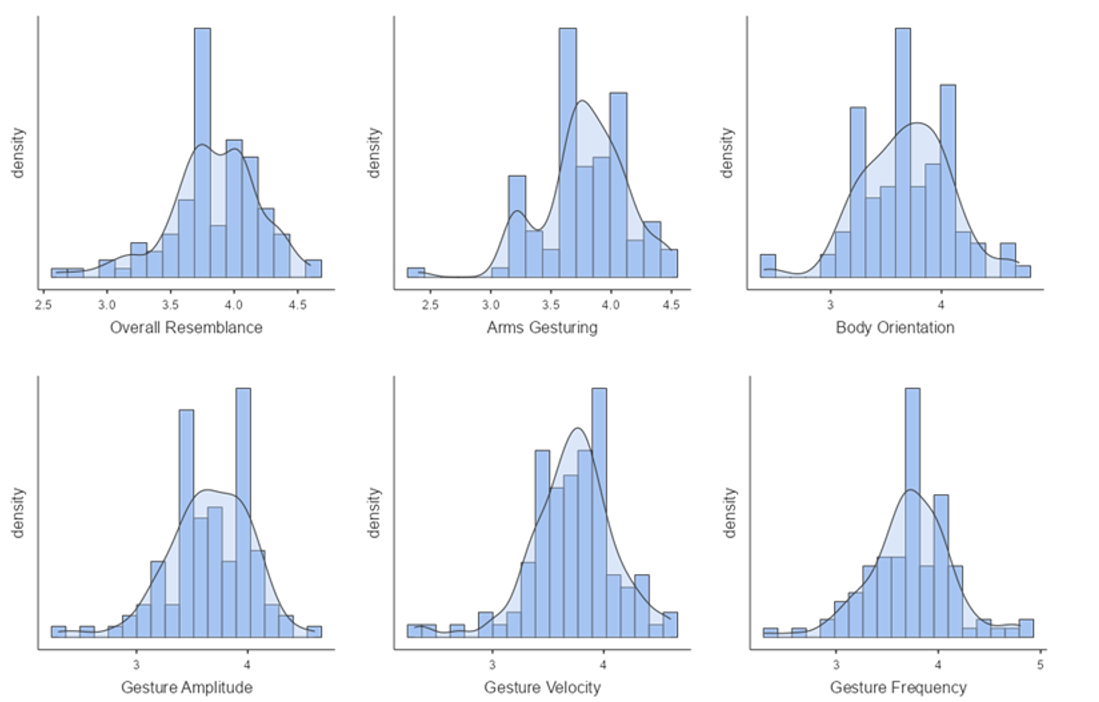}
\centering
\caption{\textbf{\textit{Body Orientation}}, \textbf{\textit{Gesture Amplitude}}, \textbf{\textit{Gesture Frequency}}, \textbf{\textit{Gesture Velocity}} for the \textit{Unseen Speakers} condition}
\label{6factors_unseen}
\end{figure}

\textbf{Study 2 - \textit{Unseen Speakers.} }
\begin{figure}[H]
\includegraphics[width=11cm]{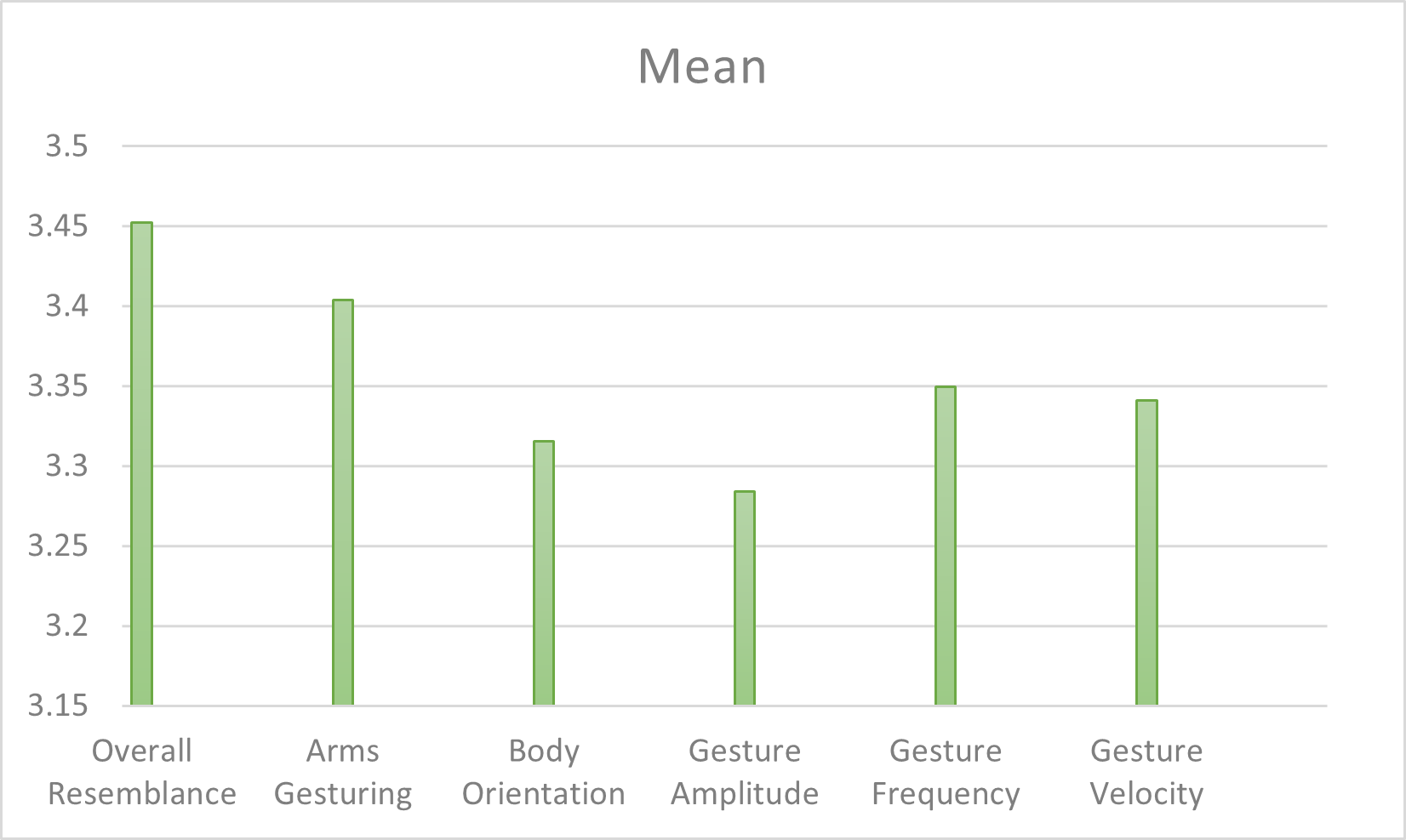}
\centering
\caption{The mean scores of all the factors for \textit{Unseen Speakers} condition}
\label{results_unseen}
\end{figure}
Our second perceptive study (Study 2) aims to evaluate the style transfer of speakers \textit{unseen} during training. Figure \ref{results_unseen} illustrates the mean scores obtained on the 6 factors for the condition \textit{"unseen speakers"}. On a \textit{5 likert scale}, the \textbf{\textit{overall resemblance}} factor obtained a score of 3.45, which means that there is an overall resemblance between \textbf{ZS-MSTM}'s 2D animations and the \textit{unseen target style}. The resemblance is also reflected by the mean scores of \textbf{\textit{arms gesturing}}, \textbf{\textit{body orientation}}, \textbf{\textit{gesture amplitude}}, \textbf{\textit{gesture frequency}}, as well as \textbf{\textit{gesture velocity}}, which is between 3.28 and 3.41. We observed that for all factors, most of the participants gave a score between 3 and 4, as depicted in Figure \ref{6factors_unseen}.

We additionally performed post-hoc paired samples t-tests between all the factors, and found significant results between \textbf{\textit{overall resemblance}} and all the other factors (p $\leq$ 0.014). 

\begin{figure}[h!]
\includegraphics[scale=0.7]{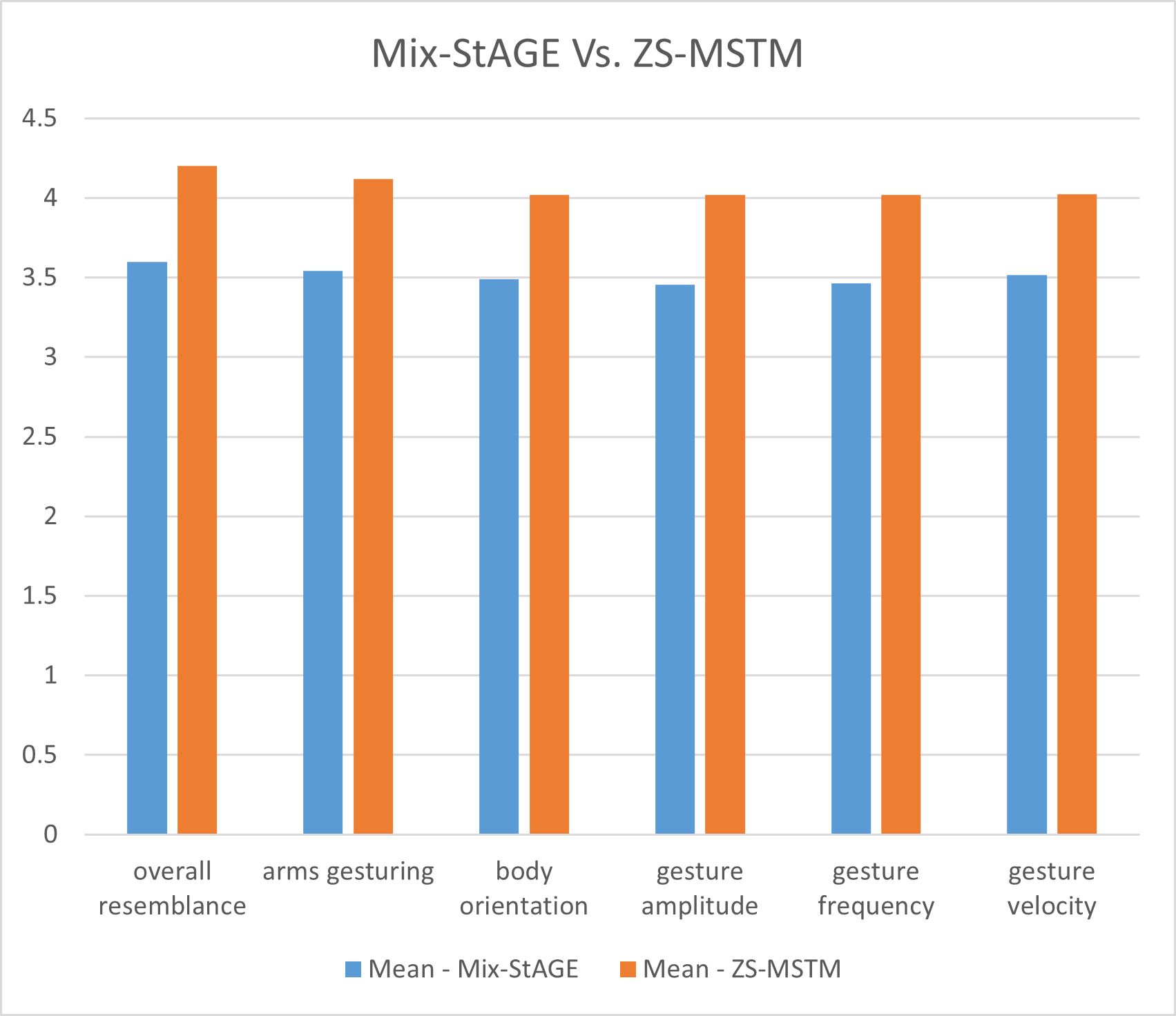}
\centering
\caption{\textbf{\textit{ZS-MSTM}} Vs. \textbf{\textit{Mix-StAGE}}}
\label{results_comparison_mixstage}
\end{figure}

\textbf{Study 3 - Comparing with Mix-StAGE. }The third perceptive study aims to compare the performance of our model with respect to the State of the Art, \textit{Mix-StAGE}. Figure \ref{results_comparison_mixstage} illustrates the mean scores obtained for the two conditions \textit{Mix-StAGE} and \textit{ZS-MSTM}, w.r.t the 6 factors.
As shown in Figure \ref{results_comparison_mixstage}, for all the factors, our model obtained higher mean scores than \textit{Mix-StAGE}. Our model performs better than \textit{Mix-StAGE} in terms of the \textbf{\textit{overall resemblance}} of the generated gestures w.r.t the animations produced with the \textit{target style} (mean score \textit{ZS-MSTM} (4.2) $\geq$ mean score \textit{Mix-StAGE} (3.6)). More specifically, the resemblance between the synthesized 2D gestures of \textit{ZS-MSTM} and the target style is greater than the one between \textit{Mix-StAGE} and the target style. This result is also reflected in the resemblance of the \textit{\textbf{arms gesturing}}, \textbf{\textit{body orientation}}, \textbf{\textit{gesture amplitude}}, \textbf{\textit{gesture frequency}} and \textbf{\textit{gesture velocity}} of our model's produced gestures w.r.t the \textit{target style}. More specifically, our model obtained a mean score between 4 and 4.2 for all the factors, while \textit{Mix-StAGE} obtained a mean score between 3.8 and 3.6 for all the factors.
We additionally conducted post-hoc paired t-tests between the factors in condition \textit{Mix-StAGE} and those in \textit{ZS-MSTM}. We found significant results between all the factors in the condition \textit{Mix-StAGE} and those in \textit{ZS-MSTM} ($p<0.001$ for all). These results show that the mean scores for all the factors in condition \textit{ZS-MSTM} are significantly greater than those \textit{Mix-StAGE}.
Thus, we can conclude that our model \textit{ZS-MSTM} can successfully render animations with the style of another speaker, going beyond the state of the art \textit{Mix-StAGE}.

\section{Discussion and Conclusion}\label{discussion}
We have presented \textbf{\textit{ZS-MSTM}}, the first approach for zero-shot multimodal style transfer for 2D pose synthesis that allows the transfer of style from any speakers \textit{seen} or \textit{unseen} during the training phase. To the best of our knowledge, our approach \textbf{\textit{ZS-MSTM}} is the first to synthesize gestures from a source speaker, which are semantically-aware, speech driven and conditioned on a multimodal representation of the style of target speakers, in a zero-shot configuration i.e., without requiring any further training or fine-tuning. \textbf{\textit{ZS-MSTM}} can learn the style latent space of speakers, given their multimodal data, and independently from their identity. It can synthesize body gestures of a source speaker, given the source speaker's mel spectrogram and text semantics, with the style of another target speaker given the target speaker's multimodal behavior style that is encoded through the mel spectrogram, text semantics, and pose modalities. Moreover, our approach is \textit{zero-shot}, thus is capable of transferring the style of unseen speakers. It is not not limited to \textit{PATS} speakers, and can produce gesture in the style of any newly coming speaker without further training or fine-tuning, rendering our approaches \textit{zero-shot}. \textit{Behavioral style} is modelled based on multimodal speakers' data, and is \textit{independent} from the \textit{speaker's identity }("ID"), which allows our model to generalize style to new \textit{unseen} speakers. We validated our approach by conducting objective and subjective evaluations. The results of these studies showed that \textbf{\textit{ZS-MSTM}} generates stylized animations that are close to the target style, for target speakers that are \textit{seen} and \textit{unseen} by our model. \textcolor{black}{The results of our ablation studies (see Table \ref{tab:ablation}) suggest that all three modalities (text, audio, and pose) are important for the performance of our \textit{ZS-MSTM} model in style transfer tasks. When any one of these modalities is removed from the style encoder, the \textit{L1 distance} between the model's predictions and the ground truth increases, indicating lower performance. This shows the importance of incorporating multiple modalities for better style transfer in our model. }Moreover, we compared the performance of \textbf{\textit{ZS-MSTM}} w.r.t the state of the art \textbf{\textit{Mix-StAGE}} and results showed that \textbf{\textit{ZS-MSTM}} performs better in terms of \textit{overall resemblance} of the generated gestures w.r.t the animations produced with the \textit{target style}. \textbf{\textit{ZS-MSTM}} can generalize style to new speakers without any fine-tuning or additional training, unlike \textit{Mix-StAGE}. Its independence from the speaker's identity "ID" allows the generalization without being constrained and limited to the speakers used for training the model. DiffGAN was later on proposed by \cite{Ahuja_CVPR_lowRes} as an extension to \textit{Mix-StAGE}, and an approach that performs \textit{few-shot} style transfer strategy based on neural domain adaptation accounting for cross-modal grounding shift between source speaker and target style. However this adaptation still requires 2 minutes of the style to be transferred which is not required by our model. \textcolor{black}{Our comparison with the baseline \textit{DiffGAN} model shows that our \textbf{\textit{ZS-MSTM}} model outperforms it in terms of both \textit{L1 distance} and \textit{PCK} metrics. This shows that our model is better at generating accurate human poses, especially when transferring styles that it has not seen during training. Overall, our results suggest that our \textit{ZS-MSTM} model is a promising approach for style transfer tasks in human pose estimation, as it can leverage multiple modalities to generate poses that are accurate.}

Our approach allows the transfer of style from any speakers \textit{seen} or \textit{unseen} during the training phase. \textit{behavior style} was never viewed as being \textit{multimodal}; previous works limit behavior style to arm gestures only. However, both \textit{text} and \textit{speech} convey \textit{style} information, and the embedding vector of \textit{style} must consider the three modalities. Our assumption was confirmed by our post-hoc t-SNE analysis of the distributions of the style vectors at the output of each modality. We found that the motion style depends mainly on the body \textit{pose modality}, followed by the \textit{speech modality}, then the \textit{text semantics modality}. We conducted an objective evaluation and three perceptive studies. The results of these studies show that our model produces stylized animations that are close to the target speakers style even for \textit{unseen} speakers. 

While we have made some strides, there are still some limitations. The main limitation of \textit{\textbf{ZS-MSTM}} is that it was not evaluated on an ECAs. The main reason is that it was trained on the \textit{PATS Corpus}, which include 2D poses. The graphical representation of the data as 2D stick figure is not always readable, even when being projected on the video of a human speaker. The main reason behind this problem is that the animation is missing information on the body pose in the Z direction (the depth axis). An interesting direction for future work is to extend our model to capture the different gesture shapes and motion. Gesture shapes convey different meanings. For example, a pointing index can indicate a direction. Hand shapes and arm movement can describe an object, an action, etc. Several attempts have looked at modelling metaphoric gestures (\cite{ravenet2018automating}), or iconic gestures (\cite{bergmann2009gnetic}). Most generative models of gestures do not compute the gesture shapes and motions for those specific gesture types. Extending our model to capture gesture shapes and motion would require extending the Corpora PATS, to include specific annotations related to gestures shapes and to identify better representations (such as image schemas \cite{grady2005image} for metaphoric gestures).

\section{Acknowledgment}
This work was performed within the Labex SMART (ANR-11-LABX-65) supported by French state funds managed by the ANR within the Investissements d’Avenir programme under reference ANR-11-IDEX-0004-02.

\bibliographystyle{unsrtnat}
\bibliography{references} 

\end{document}